

\magnification=\magstephalf
\hsize=13cm
\vsize=20cm
\overfullrule 0pt
\baselineskip=13pt plus1pt minus1pt
\lineskip=3.5pt plus1pt minus1pt
\lineskiplimit=3.5pt
\parskip=4pt plus1pt minus4pt

\def\negenspace{\kern-1.1em}


\newcount\secno
\secno=0
\newcount\susecno
\newcount\fmno\def\z{\global\advance\fmno by 1 \the\secno.
                       \the\susecno.\the\fmno}
\def\section#1{\global\advance\secno by 1
                \susecno=0 \fmno=0
                \centerline{\bf \the\secno. #1}\par}
\def\subsection#1{\medbreak\global\advance\susecno by 1
                  \fmno=0
       \noindent{\the\secno.\the\susecno. {\it #1}}\noindent}

\def\sqr#1#2{{\vcenter{\hrule height.#2pt\hbox{\vrule width.#2pt
height#1pt \kern#1pt \vrule width.#2pt}\hrule height.#2pt}}}

\newcount\refno
\refno=1
\def\y{\the\refno}
\def\myfoot#1{\footnote{$^{(\y)}$}{#1}
                 \advance\refno by 1}

\def\newref{\vskip 0.1pc 
            \hangindent=2pc
            \hangafter=1
            \noindent}
\def\neq{\hbox{$\,$=\kern-6.5pt /$\,$}}


\def\semidirect{\;{\rlap{$\subset$}\times}\;}



\newcount\secno
\secno=0
\newcount\fmno\def\z{\global\advance\fmno by 1 \the\secno.
                       \the\fmno}
\def\sectio#1{\medbreak\global\advance\secno by 1
                  \fmno=0
       \noindent{\the\secno. {\it #1}}\noindent}

\medskip

\bigskip\bigskip
\centerline{\bf YANG--MILLS CONFIGURATIONS FROM}

\centerline{\bf 3D RIEMANN--CARTAN GEOMETRY}
\bigskip

\centerline{by}
\bigskip
\centerline{{Eckehard W. Mielke, Yuri N. Obukhov$^{\diamond}$,
and Friedrich W. Hehl}}
\bigskip
\bigskip
\centerline{Institute for Theoretical Physics, University of
Cologne}
\centerline{D--50923 K\"oln, Germany}

\bigskip\bigskip
\bigskip
\centerline{\bf Abstract}

Recently, the {\it spacelike} part of the $SU(2)$ Yang--Mills equations
has been identified with geometrical objects of a three--dimensional
space of constant Riemann--Cartan curvature. We give a concise
derivation of this Ashtekar type (``inverse Kaluza--Klein")
{\it mapping} by employing a $(3+1)$--decomposition of
{\it Clifford algebra}--valued torsion and curvature two--forms. In the
subcase of a mapping to purely axial 3D torsion, the corresponding
Lagrangian consists of the
translational and Lorentz {\it Chern--Simons term} plus cosmological term
and is therefore of purely topological origin.
\bigskip
\vfill

\noindent $^{\diamond})$Alexander von Humboldt fellow. Permanent address:
Department of Theoretical Physics, Moscow State University, 117234
Moscow, Russia

\noindent Email:
pke27@rz.uni-kiel.d400.de , yo@thp.uni-koeln.de , hehl@thp.uni-koeln.de
\eject

\sectio{\bf Introduction}
\medskip
The formalism to be developed below describes a certain mapping
between Yang--Mills configurations and a suitable Riemann--Cartan
space, that is, a Riemannian space with a metric--compatible
connection --- and thus with a torsion. There is no unique way of
constructing such a mapping, and a number of recent papers were
devoted to the discussion of its definition and properties [1-10]. The
case of the $SU(2)$ group was treated first [1-8], moreover, there are
attempts to generalize these approaches to $SU(3)$ [9,10,7] (see also
earlier models developed in [11-12]). The results obtained so far can
be summarized as follows:

Yang--Mills field configurations on three-- and four--dimensional
manifolds generate an effective Riemann-Cartan (in certain models,
Riemannian) geometry on a space (or spacetime), and vice versa, the
Riemann-Cartan geometry induces Yang--Mills gauge fields. In order to
understand the basic ideas involved and to formulate the main
differences in the existing approaches, let us look at the Yang--Mills
equations. These can formally be written {\it without} the use of any
metric on an arbitrary smooth manifold $M$, see eqs.(2.2) below. The
metric comes into play only when the constitutive relations are
enforced between the Yang--Mills field strength $F$ and the excitation
$H$ (generalizing the terminology of standard Maxwell theory), such as
$^\ast F = H$ in the case of a quadratic Yang--Mills Lagrangian.  The
Hodge operator $^\ast$ is defined by some metric on $M$. In general,
this metric is unrelated to the Yang--Mills field. A well known
example is the Yang--Mills theory in Minkowski spacetime: The metric
is fixed and does not feel dynamics of the Yang--Mills (or gauge)
field. However, after the mapping the gauge field itself induces an
``effective metric'' on $M$ and thereby an Hodge operator on $M$,
which we will denote by $^\#$. In general, the $^\ast$ and the $^\#$
are different operators.

In solving the Yang--Mills equations one always needs some knowledge
about the geometry of $M$. If an independent metric exists, then one
writes $H=\,^\ast F$ (with respect to this metric), and finally one
has to solve the differential equations and to find a gauge field
configuration {\it on a prescribed background metric}.

The case $^\ast=\,^\#$, which will be developed below, is somewhat
different. One starts from a Yang--Mills field configuration $F$,
formally constructs an ``effective geometry" on $M$, and eventually it
remains to find the geometrical variables in such a way that the
vacuum Yang--Mills equation $D^\# F=0$ is satisfied automatically.
This is what will be done below. One could call this approach
``Yang--Mills equations without Yang--Mills equations". More
specifically, using the formalism of Clifford--valued forms, we will
start from 3--dimensional (or $3D$) Einstein(--Cartan) type equations
with cosmological constant, (4.3). The geometry is described in terms
of the frame and the torsion together with curvature two--forms. Via
(4.4)--(4.5), these define the Yang--Mills gauge field. The
Yang--Mills equations {\it on this effective geometry} are recovered
from the Bianchi identities for curvature and torsion.  In general,
going in the opposite direction, from the Yang--Mills theory to the
Riemann--Cartan geometry, one could have richer structures, including
the case $^\ast\neq\,^\#$, the latter of which is discussed in [8],
for instance.
\bigskip
\goodbreak
\sectio{\bf  The $(3 + 1)$--decomposition of the Yang--Mills equations}

In terms of $SU(2)$ generators $\tau_a:=-(i/2)\sigma_a$, where $a,b,
\cdots =1,2,3$ and the $\sigma_a$ are the Pauli matrices, the $SU(2)$
gauge connection $A$ and the Yang--Mills field strength $F$ read,
respectively:
$$ A:=A_i^{a}\,\tau_a\, dx^i, \qquad F:=dA + A\wedge A=
{1\over 2}F_{ij}^a\, \tau_a\, dx^i\wedge dx^j \, . \eqno(\z)$$
Here $i,j,\cdots=0,1,2,3$ are (holonomic) coordinate indices of the
underlying smooth manifold $M$.

With $\check D:=d+A$ as the $SU(2)$--gauge covariant exterior
derivative, the Yang--Mills equations are $$\check{D} F\equiv
0,\qquad\qquad\check{D}H =J \, , \eqno(\z)$$ where the excitation
two--form $H=H^a\, \tau_a$ and the external current three--form
$J=J^a\, \tau_a$ are $su(2)$--valued. We remind ourselves of the
definition $H=-2\partial L_{YM}/\partial F$ of the excitation in terms
of the Lagrangian four--form $L_{YM}$.

We assume that the four--dimensional manifold $M$ admits a foliation
into three--dimensional hypersurfaces $\Sigma_t$, where $t$ denotes a
monotonically increasing parameter. By means of a vector field $${n}
:= \partial_{t} -N^{A}\partial_{A} =N\,\widetilde n.\eqno(\z)$$ (which
will turn out to be time--like later on and) which is normal to a
hypersurfaces $\Sigma_t$, we can perform $(3+1)$--decompositions of
the Yang--Mills equations.  Here $A,B,\cdots =1,2,3$ are spatial
indices. Compared to other notations, it is rescaled by the lapse
function $N$, see Ref.[13] for details.  With respect to this normal
vector field ${n}$, we define, for a $p$--form $\Psi$ and {\it
arbitrary} lapse $N$ and shift $N^{A}$, the {\it normal} part by
$$^{\bot}\Psi := dt\wedge\Psi_{\bot};\qquad\qquad
\Psi_{\bot}:= {n}\rfloor\Psi,\eqno(\z)$$
and the part {\it tangential} to the hypersurface $\Sigma_{t}$ by
$$\underline{\Psi} := {n}\rfloor (dt\wedge\Psi) =
(1\; -\; ^{\bot})\Psi ,\qquad\qquad
{n}\rfloor\, \underline{\Psi}\equiv 0 .\eqno(\z)$$
These  projection operators ``$\bot$" and
``$\underline{\;\;}$" form a complete set, see the Appendix for further
rules of calculation.

As in Maxwell's theory, the Yang--Mills field strength $F$ decomposes
into a ``magnetic''and an ``electric'' piece, respectively:
$$ B:=\underline{F} = \underline{d}\underline{A} +
\underline{A}\wedge\underline{A}=B^a\tau_a\, ,\qquad
E:= -F_{\bot} = \underline{\check D} A_{\bot}-
\ell_{{n}}\underline{A} =E^a\tau_a\,. \eqno(\z)$$

With these abbreviations, the homogeneous Yang--Mills equation
$\check{D}F\equiv 0$, which is, in fact, a Bianchi type identity,
decomposes as follows into constraint and propagation equation:
$$\underline{\check D}B= 0 \, ,\qquad\qquad\underline{\check D} E +
\check{ \L}_n B =0\, .\eqno(\z)$$
Analogously, the $(3+1)$--decomposition
of the inhomogeneous equation $\check{D}H=J$ yields constraint
and propagation equation:
$$ \underline{\check D}\,{\cal D} =\underline{J} =:\rho
\, ,\qquad\qquad\underline{\check D}{\cal H}-
\check{ \L}_n{\cal D} = -J_{\bot}=: j
\,. \eqno(\z)$$
Due to the specification of the charge density three--form $\rho$
and the current two--form $j$, we can identify
$$ {\cal D} := \underline{H}\, ,
\qquad {\rm and} \qquad{\cal H} := H_{\bot}= n\rfloor H \eqno(\z)$$
as the Yang--Mills analogues of the electric excitation two--form
and the magnetic excitation one--form,
respectively, cf. Hehl [14] for details in the Maxwellian $U(1)$--case.
An overview of the eqs.(2.6)--(2.9), together with their number
of independent components, is given in Table 1.
\bigskip

\noindent{\bf Table 1.} $SU(2)$ Yang--Mills field variables and equations.
We display the number of independent components in $(3+1)$ dimensions
before and after the decomposition in space and time.
$$\offinterlineskip\tabskip=0pt\vbox{\halign to 1.0\hsize
{\strut\vrule width0.8pt\quad#\tabskip=0pt plus 100pt
& #\hfill\quad
&\vrule#&
&\quad #\quad
&\vrule#
&\quad #\quad
&\vrule#
&\quad #\quad
&\vrule#
&\quad #\hfill\quad
 \tabskip=0pt
&\vrule width0.80pt#
\cr
\noalign{\hrule}\noalign{\hrule}
&\bf variables and
  &&\multispan 5\hfill
components in $(3+1)$ dimensions\hfill&&remarks\hfill&\cr
&\bf field equations &&total && tangential&&normal&& &\cr
\noalign{\hrule}\noalign{\hrule}
&potential       $A$ && 12 && \bf 9 && 3&&                &\cr
&field strength $F$  && 18 && \bf 9 && 9&& $B$ \& $E$ &\cr
&$\check{D}F=0$&& 12 && \bf 3 && 9&& constr. \& propag.   &\cr
&excitation $H$      && 18 &&\bf 9 && 9&& $\cal D$ \& $\cal H$ &\cr
&$\check{D}H=J$ && 12 && \bf 3 && 9&& constr. \& propag.  &\cr
\noalign{\hrule\hrule} }}$$

\bigskip\goodbreak
\sectio{\bf Clifford algebra--valued field variables in $4D$
Riemann--Cartan geometry}
\medskip
The group $SU(2)$ is isomorphic to (the covering group of) the
rotation group $SO(3)$. If one interpretes the $SO(3)$ as acting in
ordinary three--dimensional Euclidean space, then the $(3+3)$
parameter Euclidean group $SO(3)\semidirect R^3$ --- with $R^3$ as
translations --- represents the group of motion. The {\it gauging} of
this external group yields a $3D$ {\bf R}iemann--{\bf C}artan geometry
with curvature and torsion. Accordingly, this geometry emerges in a
very natural way in the context of a gauge procedure. In order to link
the $3D$ RC--geometry to the four--dimensional manifold $M$ discussed
in Sec.1, it is plausible to start with a $4D$ RC--geometry instead
and to apply again a $(3+1)$ decomposition to the corresponding
expressions so as to arrive at a $3D$ RC--geometry in the end.

One can describe the $4D$ RC--geometry by means of a very elegant
formalism, namely by employing Clifford algebra--valued exterior
differential forms, see Hehl et al.[15].  To this end we use the Dirac
matrices $\gamma_{\alpha}$ in the Bjorken and Drell convention [16]
with signature $(+---)$. They obey the anticommutation relations
$$\gamma_\alpha\gamma_\beta+\gamma_\beta\gamma_\alpha=2o_{\alpha\beta}
\,  1_4\,,\eqno(\z)$$
where $\alpha, \beta,\cdots=$\^0, \^1, \^2, \^3 denote the
(anholonomic) indices of the frame field $e_\alpha$ which is assumed
to be {\it orthonormal}.  The 16 matrices
$\{1_4,\gamma_\alpha,\sigma_{\alpha\beta},\gamma_5,
\gamma_5\gamma_\alpha\}$ form a basis of a
{\it Clifford algebra} in four dimensions.

Following the notation of Hehl et al.[15], see Refs.[17, 18]
for earlier approaches to ``color geometrodynamics" and Ref.[19] for
Clifford bundles, the constant $\gamma_\alpha$ matrices can be
converted into Clifford algebra--valued one-- or three--forms, respectively:
$$\gamma:=\gamma_\alpha\vartheta^\alpha\,,\qquad\;^{*}\gamma=\gamma^\alpha
  \eta_\alpha= \gamma_5\;^*(\gamma\wedge\gamma\wedge\gamma)\,.\eqno(\z)$$
Here $\eta$ is the volume four--form and $\eta_\alpha:=e_\alpha\rfloor\eta=
\;^{*}\vartheta_\alpha$ is the
coframe ``density''. The totaly antisymmetric
product of Dirac matrices has  now be identified with the zero--form
$\gamma_5:= (i/4!)\;^*(\gamma\wedge\gamma\wedge\gamma\wedge\gamma)$ .
Another useful element of the Clifford algebra is the Lorentz generator
$$\sigma_{\alpha\beta}:={i\over 2}
  (\gamma_\alpha\gamma_\beta-\gamma_\beta\gamma_\alpha)\,,\qquad
  [\gamma_\alpha,\sigma_{\beta\gamma}]=
  2i(o_{\alpha\beta}\,\gamma_\gamma-o_{\alpha\gamma}\,\gamma_\beta)\,.
  \eqno(\z)$$
and its {\it Lie} (or right) dual
$$\sigma^{\star}_{\alpha\beta}:=
\sigma^{\gamma\delta} \,{1\over 2}\eta_{\alpha\beta\gamma\delta}
 = i\gamma_5 \,\sigma_{\alpha\beta}\, ,  \eqno(\z)$$
where the components of the metric volume element four--form
read $\eta_{\hat{0}\hat{1}\hat{2}\hat{3}}= +1$.
The associated two-forms are given by
$$\sigma:={1\over 2}\sigma_{\alpha\beta}\,\vartheta^\alpha\wedge
  \vartheta^\beta={i\over 2}\,\gamma\wedge\gamma\,, \qquad
\,^*\sigma={1\over 2}\sigma_{\alpha\beta}\,\eta^{\alpha\beta}
=:\sigma^{\star}\, .\eqno(\z)$$

In terms of the Clifford algebra--valued connection and the
$\overline{SO}(1,3)\cong SL(2,C)$--covariant
exterior derivative
$$\Gamma:= {i\over 4}\Gamma^{\alpha\beta}\,\sigma_{\alpha\beta}\; ,
\qquad D=d+[\Gamma,\quad]\; ,\eqno(\z)$$
respectively, Clifford algebra--valued {\it torsion} and {\it curvature}
two--forms
can be defined as follows:
$$\Theta :=D\gamma =T^{\alpha}\gamma_{\alpha}\;, \qquad
\Omega := d\Gamma +\Gamma\wedge \Gamma =
{i\over 4}R^{\alpha\beta}\,\sigma_{\alpha\beta}\, .\eqno(\z)$$
The algebra--valued form commutator is given by
$[\Psi, \Phi] := \Psi\wedge\Phi - (-1)^{pq} \Phi\wedge\Psi$, cf.
Ref.[18], p.26. Then the two Bianchi identities
$$DT^{\alpha}\equiv R_{\beta}{}^{\alpha}\wedge\vartheta^{\beta}\;,
\qquad \qquad
DR^{\alpha\beta}\equiv 0\, \eqno(\z)$$
of RC--spacetime take the rather simple form
$$ D\Theta \equiv [\Omega ,\gamma]\, ,\qquad\qquad
D\Omega\equiv 0 \, .\eqno(\z)$$
These structures of a RC--space are collected in Table 2.
\bigskip
\noindent{\bf Table 2.} Variables and Bianchi identities
in Riemann--Cartan space. The number of independent components is
displayed in $(3+1)$ dimensions before and after the decomposition in
space and time. The numbers in boldface count the independent
components of those expressions which
are involved in the mapping procedure to the tangential Yang--Mills
expressions of Table 1.
$$\offinterlineskip\tabskip=0pt\vbox{\halign to 1.0\hsize
{\strut\vrule width0.8pt\quad#\tabskip=0pt plus 100pt
&#\hfill\quad
&\vrule#&
&\quad #\quad
&\vrule#
&\quad #\quad
&\vrule#
&\quad #\quad
 \tabskip=0pt
&\vrule width0.80pt#
\cr
\noalign{\hrule}\noalign{\hrule}
&\bf variables and
  &&\multispan 5\hfill
components in $(3+1)$ dimensions \hfill&\cr
&\bf Bianchi identities &&total && tangential&&normal&\cr
\noalign{\hrule}\noalign{\hrule}
&coframe $\gamma$     && 16 && 9+3 &&  3+1 &\cr
&connection $\Gamma$     && 24 && {\bf 9}+9 &&  3+3 &\cr
&curvature $\Omega$           && 36 && {\bf 9}+9 &&  3+3 &\cr
&2nd Bianchi $D\Omega=0$                  && 24 && {\bf 3}+3 &&  9+9 &\cr
&torsion $\Theta$             && 24 && {\bf 9}+3 &&  9+3 &\cr
&1st Bianchi $D\Theta=[\Omega,\gamma]$  && 16 && {\bf 3}+1 &&  9+3 &\cr
\noalign{\hrule\hrule} }}$$
\bigskip
For a possible mapping of the vacuum Yang--Mills equations to the
Bianchi identities, see Tables 1 and 2, it is tempting to require the
consistency condition $$[\Omega , \gamma ]=0\,,\eqno(\z)$$ i.e.\ to
admit only a Weyl, a symmetric tracefree Ricci, and a
scalar piece in the curvature, cf.[20]. By
differentiation we find $$
[\Omega ,\Theta ]=0\, .\eqno(\z)$$
Because of the validity of the 2nd Bianchi identity $D\Omega\equiv 0$, we
try the ansatz of constant curvature, $$\Omega = {\Lambda\over 4}\,
\gamma\wedge\gamma =- {{i\Lambda}\over 2}\, \sigma \,,\eqno(\z)$$ i.e.,
the curvature scalar is the only surviving piece of the curvature.
Therefore (3.12) fulfills (3.10). Moreover, for $\Lambda\neq 0$, the
2nd Bianchi identity implies $$[\Theta ,\gamma] =0 \,, \eqno(\z)$$
which, in {\it four} dimensions, amounts to 24 independent conditions.
Accordingly (3.13) yields vanishing torsion [21].  Thus we have to
abandon, or rather to weaken, the degenerate four--dimensional ansatz
(3.12).

\goodbreak\bigskip
\sectio{\bf Mapping Yang--Mills fields into
3D Riemann--Cartan geometry}
\bigskip
As a weaker consistency condition, we require merely the tangential
part of (3.10) to hold: $$ [\underline{\Omega} , \underline{\gamma}
]=0\, .\eqno(\z)$$ By differentiation we find, instead of (3.11),
$$[\underline{\Omega} , \underline{\Theta} ]=0\,.\eqno(\z)$$

This time we only postulate the {\it tangential} curvature to be constant
as one of the key relations of our work:
$$\underline{\Omega} = {\Lambda\over 4}\,
\underline{\gamma}\wedge\underline{\gamma} =
- {{i\Lambda}\over 2}\, \underline{\sigma} \,.\eqno(\z)$$ It fulfills
(4.1). Thus $$\underline{D}\;\underline{\Theta}=0\,.\eqno(\z)$$ For
$\Lambda\neq 0$, as a consequence of the tangential part of the 2nd
Bianchi identity, we have $$[\underline{\Theta} ,\underline{\gamma}]
=0 \, ,\eqno(\z)$$ Now we have specified the type of
$(3+1)$--dimensional RC--geometry which we want to consider, namely the
one characterized by the tangential curvature (4.3) and a tangential
torsion constrained by (4.5).

It is convenient to denote
$$\omega:={i\over 4}\,\Gamma^{AB}\sigma_{AB}\,,\qquad\varphi:=
{i\over 4}\,\Gamma^{\hat{0}A}\sigma_{\hat{0}A}\,.\eqno(\z)$$
Then the curvature decomposes as follows:
$$\underline{\Omega}=(\underline{d}\,\underline{\omega}+\underline{\omega}
\wedge\underline{\omega})+(\underline{d}\,\underline{\varphi}+
\underline{\omega}\wedge\underline{\varphi}+\underline{\varphi}\wedge
\underline{\omega})\,.\eqno(\z)$$

Without restricting the generality of our considerations, we can fix
the coframe such that $\vartheta^{\hat{0}}$ is aligned with the
$t$--axis. In this {\it time gauge} for the coframe [22],
$$\underline{\vartheta}^{\hat{0}}=0\,,\eqno(\z)$$
the relation (4.3) splits into two sets of equations:
 $$\underline{d}\,\underline{\omega}+\underline{\omega}\wedge
\underline{\omega}=-{i\Lambda\over 2}\,\underline{\sigma}\,,\eqno(\z)$$
 $$\underline{d}\,\underline{\varphi}+\underline{\omega}\wedge
\underline{\varphi}+\underline{\varphi}\wedge\underline{\omega}=0\,.\eqno(\z)$$

We take the exterior derivative of (4.10) and substitute (4.9) and (4.10):
$$
[\underline{\varphi}\,,\underline{\sigma}]=0\,.\eqno(\z)
$$
Eq.(4.11) has only the trivial solution
$$
\underline{\varphi}=0\,.\eqno(\z)
$$
Hence we conclude that, in the time gauge (4.8), the postulate (4.3) yields
$$
\underline{\Gamma}=\underline{\omega}\,.\eqno(\z)
$$
In particular, for the torsion we find
$$
\underline{T}^{\hat{0}}=\underline{D}\,
\underline{\vartheta}^{\hat{0}}=\underline{d}\,
\underline{\vartheta}^{\hat{0}}=0\quad {\rm and}\quad \underline{D}\,
\underline{T}^{\hat{0}}=\underline{d}\,\underline{d}\,
\underline{\vartheta}^{\hat{0}}=0\,.\eqno(\z)
$$

Eq.(4.9) -- and thus (4.3) in the time gauge (4.7) -- is equivalent to the 3D
Einstein equations with cosmological term. Exactly such a 3D
Riemann--Cartan space follows also from the $(3+1)$--decomposition of
the modified double duality ansatz $\Omega =\,^*\Omega^{\star} -
i(\Lambda/2)\,\sigma$ in the gauge of purely ``magnetic" RC curvature,
see Ref.[23], Sec.6.

Let us compare our results with the tangential column in Table 2:
Because of (4.8), for the coframe 3 independent components are
cancelled and 9 are left over, likewise for the connection, see (4.6)
and (4.12), 9 are killed and 9 survive. According to (4.10), in the
time gauge, half of the curvature components vanish, that is, 9 are
only allowed for. Note, however, that in the end (4.3) admits only the
1 component of the curvature scalar as a lone surviver. The torsion
constraint (4.5), in the time gauge, kicks out the 3 components of the
axial piece of the torsion. Finally, the 1st Bianchi identity is made
to shrink by means of (4.1) and (4.14). Thus in Table 2 only the
column with the boldface numbers are left over, whereas the timelike
components in the tangential column are suppressed. This is a
necessary condition that the mapping, see Table 1, can be implemented.


\bigskip\noindent{\bf Table 3.} Mapping $3D$ Riemann-Cartan geometry to
$SU(2)$ Yang-Mills field configurations.
\bigskip
{\offinterlineskip\tabskip=0pt
\halign{\strut
\vrule#&  
\quad #\hfil& 
\vrule#&  
\quad #$\>$\hfil& 
\vrule#& 
 #& 
\vrule#& 
$\>\>$#\hfil&  
\vrule#& 
\quad #\hfil& 
\vrule#& 
\quad #\hfil &  
\vrule#  \cr   
\noalign{\hrule}
&$SU(2)$ YM in $4D$ && $\underline{\rm YM}$&&  && $3D$ RC  && RC--notions
&& comp. &\cr
\noalign{\hrule}
&potential $A$       && $\underline{A}$
 &\omit & $\;\Leftrightarrow$ &\omit & $\underline{\Gamma}$      && connection
&& 9&\cr
&field strength $F$  && $\underline{F}=B$
 &\omit & $\;\Leftrightarrow$ &\omit & $\underline{\Omega}$           &&
curvature   && 9&\cr
&$\check{D}F=0$      && $\underline{\check{D}}\,B=0$
 &\omit & $\;\Leftrightarrow$ &\omit & $\underline{D}\,\underline{\Omega}=0$
&& 2nd Bianchi
&& 3&\cr
&excitation $H$      && $\underline{H}={\cal D}$
&\omit & $\;\Leftrightarrow$ &\omit & $\underline{\Theta}$           && torsion
    && 9&\cr
&$\check{D}H=0$      && $\underline{\check{D}}\,{\cal D}=0$
&\omit & $\;\Leftrightarrow$ &\omit & $\underline{D}\,\underline{\Theta}=0$
&& 1st Bianchi with constr. &&3&\cr
\noalign{\hrule}}}
\bigskip

A comparison of Table 1 and 2 makes it obvious that we should
identify the gauge potential with the three--dimensional
RC--connection as follows: $$ \underline{A}\otimes
1_2=\underline{\Gamma}\,. \eqno(\z)$$ Under the
conditions (4.3), we can now identically satisfy the tangential
Yang--Mills equations (2.7a), (2.8a) by means of the tangential Bianchi
identities of RC--space, provided we make the following
identifications: $$B\otimes 1_2= \underline{F}\otimes 1_2
=\underline{\Omega} = -{{i\Lambda}\over 2}
\underline{ \sigma} =\Lambda\underline{\eta}^a \tau_a \otimes 1_2\,,\eqno(\z)$$
$$
{\cal D}\otimes 1_2=\underline{H}\otimes 1_2
=i\gamma_5\gamma_0\underline{\Theta}.\eqno(\z)
$$
Here we have used the relation (5.10) of the Appendix, which holds
for tetrads in the time--gauge. (In order to recover $\underline{H} =
\underline{H}^a\, \tau_a $, the  factor
$\gamma_5\gamma_0 =
\pmatrix{0& -1\cr 1& 0\cr}$ is needed in order to shift the
$SU(2)$ generators $\tau_a$ in
$\gamma_a$ to the main diagonal.)

Since $H=\,^{*}F$ in the standard Yang--Mills case with quadratic Lagrangian,
its (3+1)--decomposition automatically provides a geometrical relation for the
normal parts of the Yang--Mills fields:
$$
E\otimes 1_2=-N \,^{ \underline{*}}  \underline{H}\otimes 1_2 =
-i N \gamma_5\gamma_0\,^{ \underline{*}}
\underline{\Theta},\eqno(\z)
$$
$$
{\cal H}\otimes 1_2= N \,^{ \underline{*}}\underline{F}\otimes 1_2
=-N\Lambda\underline{\vartheta}^{a }\tau_a\otimes 1_2 . \eqno(\z)
$$

The identification (4.16) implies that the 3D coframe is given by
$$
\underline{\vartheta}^{a } =- {1\over \Lambda}\,^{ \underline{*}}B^a \eqno(\z)
$$
In the time gauge, the three--metric can now be expressed via (4.20) as
$$\underline{g} =o_{ab} \underline{\vartheta}^{a }\,
\underline{\vartheta}^{b } = -\delta_{ab}
{1\over{ \Lambda^2}}\,^{ \underline{*}}B^a\;^{ \underline{*}}B^b
=  -{1\over{4 \Lambda^2}}tr\, (\,^{ \underline{*}}\underline{\Omega}
 \,^{ \underline{*}}\underline{\Omega})\, . \eqno(\z)$$
Observe, that the Hodge star $\,^{ \underline{*}}$ is implicitly
depending on the Yang--Mills field strength $B$. This is reflected, e.g.,  in
the determinant of the  3D coframe, which is, by definition,
the volume three--form
$$\underline{\eta} = {1\over{ 3 \Lambda^2}} \,^{ \underline{*}}B^a\wedge
B_a \, . \eqno(\z)$$

The relations (4.3) and (4.20) have been found already in a component
approach [1,5].

The torsion constraint (4.5) is solved by
$$
\underline{\Theta}= \underline{t}^{(ab)}{\underline{\eta}}_{a}\gamma_{b},
\eqno(\z)
$$
where symmetric tensor $\underline{t}^{(ab)}$ satisfies, in view of (4.4),
the differential equation
$$
\underline{e}_{a}\rfloor\underline{D}\underline{t}^{(ab)}=0.\eqno(\z)
$$

\goodbreak\bigskip
\sectio{\bf Concluding remarks}
\medskip

In general, the 3D constraint (4.3) of constant RC curvature
can be derived from a Lagrangian only by imposing it via Lagrange multipliers.
However, for the subcase of purely axial torsion, which  corresponds to the
choice $\underline{t}^{(ab)}= -(2/\ell)\,o^{ab}$ with a constant $\ell$, one
can do better: In the generalization of the DJT gravity model with torsion of
Ref.[24], there occurs  the  3D Lagrangian
$$
V_{\infty}  = {\Lambda\over{4\ell}} Tr\,
( \underline{\gamma} \wedge  \;^{\underline{*}}\underline{\gamma} )
 -{\Lambda\over{8}} Tr\,
( \underline{\gamma} \wedge  \underline{\Theta} )
-{1\over2} Tr\, (  \underline{\Gamma}\wedge \underline{\Omega} -
{1\over 3} \underline{\Gamma}\wedge
 \underline{\Gamma}\wedge  \underline{\Gamma} )\, . \eqno(\z)
$$
Since it consists out of a cosmological term plus the tangential part of
the  translational and Lorentz  Chern--Simons term, it is of purely
{\it topological} origin. Variation with respect to $\underline{\gamma}$
and $\underline{\Gamma}$ yields the field equations
$$
\underline{\Theta}={2\over{\ell}}\;^{\underline{*}}\underline{\gamma}\, ,
\qquad\underline{\Omega} = -
{i\over 2}\, \Lambda\,\underline{ \sigma }\, ,\eqno(\z)
$$
which is just (4.3) for purely axial torsion. Moreover, in that case we find
from (4.16), (4.18) that not only the ``magnetic" but also the
``electric" component of the Yang--Mills field strength is proportial
to the triad, i.e.  $E= -(4N/\ell )\, \underline{\vartheta}^a\,\tau_a\,
$. Such a mapping is considered, e.g., by Lunev [1].

We have found [24] a peculiar symmetry in the equations of these 3D models,
which has the suggestive Clifford algebra correspondence
$$
\gamma_5 \underline{\gamma}\sim \ell\underline{\Gamma}\,.\eqno(\z)
$$
Then, the 3D metric can also be expressed via $ \underline{g}=(1/4) Tr\,
(\underline{\gamma}\;\underline{\gamma})$ in terms of the Yang--Mills
connection $\underline{A}= \underline{\Gamma}$. This is actually
related to the identifications proposed in Refs. [9,10,11].

Using a quaternionic representation, Dolan [25] has already earlier
derived a similar mapping in the special case of self-- or anti--selfdual
Yang--Mills configurations. In particular, he could show that then the
underlying geometry is  even a 4D Einstein space (with Euclidean
signature).

In our approach via Clifford algebra--valued forms, the mapping from
Yang--Mills to  3D Einstein--Cartan space is rather transparent and
simple geometrically. It is akin to  Ashtekar's Hamiltonian approach [26]
to the Einstein's general relativity, except for that here the traceless
3D torsion plays an essential role. In Refs. [27, 28], however,
new canonical variables for the teleparallelism equivalent of general
relativity are generated such that the {\it complexified torsion}, as the
field strength of translations, carries the gravitational degrees of
freedom in a Yang--Mills type fashion.

\goodbreak\bigskip
\sectio{\bf Appendix:  The $(3+1)$--decomposition of coframe, torsion,
curvature, and Bianchi identities}
\medskip
In order to apply these  $(3+1)$--decompositions to
field theory or geometry, the following rules
with respect exterior multiplication and the
forming of the Hodge dual are instrumental:
$$
\eqalign{\,^{\bot}(\Psi\wedge\Phi )&=\, ^{\bot}\Psi\wedge\underline{\Phi} +
\underline{\Psi}\wedge\,^{\bot}\Phi\; \cr
&= dt\wedge (\Psi_{\bot}\wedge\underline{\Phi} + (-1)^{p}
\underline{\Psi}\wedge\Phi_{\bot})\, , \quad\qquad\qquad
\underline{\Psi\wedge\Phi}= \underline{\Psi}\wedge
\underline{\Phi},\cr }   \eqno(\z)
$$
$$
^{\bot}(^{*}\Psi )=(-1)^{p}N\,dt\wedge\,^{\underline{*}}\underline{\Psi}\; ,
\qquad\qquad  \underline{\, ^{*}\Psi}=- {1\over N}
\,^{\underline{*}}\Psi_{\bot}\,.\eqno(\z)
$$
Here $\,^{\underline{*}}$ denotes the Hodge dual in three dimensions
which, in the Bjorken and Drell conventions, is an anti--involutive operator:
$$
\,^{\underline{*}}\;^{\underline{*}}\Psi=-\Psi .\eqno(\z)
$$
The exterior derivative of a $p$--form decomposes as follows:
$$
^{\bot}(d\Psi )= dt\wedge(\ell_{{n}}\underline{\Psi} -
\underline{d}\Psi_{\bot}),\qquad\qquad \underline{d\Psi} =
 \underline{d}\; \underline{\Psi}.\eqno(\z)
$$
In the formulae above, the {\it Lie derivative} of $p$--forms along a
vector field $\xi$ is defined by
$$
\ell_{\xi}\Psi := \xi\rfloor d\Psi + d(\xi\rfloor\Psi).\eqno(\z)
$$

For Lie algebra--valued $p$--forms, the {\it gauge--covariant} Lie
derivative
$${\L}_{\xi}\Psi^{\alpha \cdots} :=\xi\rfloor
D\Psi^{\alpha \cdots}+ D(\xi\rfloor \Psi^{\alpha \cdots})=
\ell_{\xi}\Psi^{\alpha\cdots} +
 (\xi\rfloor \Gamma_{\beta}{}^{\alpha})\Psi^{\beta \cdots} + \cdots
\eqno(\z)$$
is more convenient.

The coframe decomposes as follows:
$$^\bot\vartheta^{\alpha } =
{n}^{{i}}e_{{i}}{}^{\alpha }dt \quad\Leftrightarrow\quad
\vartheta^{\alpha }_{\bot} = {n}^{\alpha },\qquad
\qquad\underline{\vartheta}^{\alpha } = e_{{i}}{}^{\alpha }dx^{{i}} -
{n}^{\alpha }dt, \eqno(\z)$$
such that
$$\vartheta^{\alpha } =  \; ^\bot\vartheta^{\alpha }   +
\underline{\vartheta}^{\alpha }. \eqno(\z)$$
For the ``tetrads in the time--gauge" [22], the
space--like one--form $\underline{\vartheta}^{\hat{0}}$ is
vanishing.
Although, we do
{\it not} assume this gauge in our subsequent
decompositions, it is convenient in the $(3+1)$ decomposition of
the Lorentz generator two--form $\sigma$. We find
$$
\sigma_{\bot} ={i\over 2}[\gamma_{\bot}\, ,\underline{\gamma}]\, ,\eqno(\z)
$$
where the normal part $\gamma_{\bot}=n^\alpha\,\gamma_{\alpha}$ generalizes
$\gamma_0$ for arbitrary lapse and shift, and
$$
\underline{\sigma}=
\underline{\eta}_{c}\,( \sigma^c\otimes 1_2)\,.\eqno(\z)
$$
Here we have used the relation $\sigma_{ab}=\eta_{\hat{0}abc}(\sigma^c\otimes
1_2) =\underline{\eta}_{abc}\, (\sigma^c\otimes 1_2)$ for the Pauli
matrices $\sigma^a$.

For the connection we obtain
$$^\bot\Gamma^{\alpha \beta}  = \Gamma_{\bot}{}^{\alpha \beta}dt =
{n}^{{i}}\Gamma_{{i}}{}^{\alpha \beta}dt,
\qquad\qquad\underline{\Gamma}^{\alpha \beta } =
\Gamma_{{i}}{}^{\alpha \beta}dx^{{i}}
-\Gamma_{\bot}{}^{\alpha \beta}dt. \eqno(\z)$$
By evaluating Cartan's structure equation $T^\alpha:= D\vartheta^\alpha$,
 we find the associated decompositions of the torsion:
$$T^{\alpha }_{\bot} = \L_{{n}}\underline{\vartheta}^{\alpha } -
\underline{D}{n}^{\alpha },\qquad
\underline{T}^{\alpha } =
\underline{D}\;\underline{\vartheta}^{\alpha }.\eqno(\z)$$

For the Riemann--Cartan curvature $R^{\alpha \beta}:=
d\Gamma^{\alpha \beta} - \Gamma^{\alpha\gamma }
\wedge\Gamma_{\gamma}{}^{\beta}$ we obtain
$$R^{\alpha \beta}_{\bot} =
\ell_{{n}}\underline{\Gamma}^{\alpha \beta} -
\underline{D}\Gamma_{\bot}{}^{\alpha \beta},\qquad
\underline{R}^{\alpha \beta} =
\underline{d}\underline{\Gamma}^{\alpha \beta} -
\underline{\Gamma}^{\alpha\gamma }
\wedge\underline{\Gamma}_{\gamma}{}^{\beta}. \eqno(\z)$$

We also exhibit the $(3+1)$--decompositions of the
Bianchi identities: For the first identity (3.8a.) we find
$$
\L_{{n}}\underline{T}^{\alpha }-
\underline{D}\,T^{\alpha}_{\bot} \equiv R_{\beta}{}^{\alpha}_{\bot}
\wedge\underline{\vartheta}^{\beta} +
\underline{R}_{\beta}{}^{\alpha }\,{n}^{\beta} \eqno(\z)
$$
and
$$
\underline{D}\underline{T}^{\alpha}\equiv\underline{R}_{\beta}{}^{\alpha}
\wedge\underline{\vartheta}^{\beta }\, .\eqno(\z)
$$
The second Bianchi identity (3.8b) decomposes into
$$
\underline{D}\;\underline{D}(\Gamma_{\bot}{}^{\alpha \beta}) \equiv
\underline{R}_{\gamma }{}^{\alpha}\,\Gamma_{\bot}{}^{\gamma \beta} -
\underline{R}_{\gamma}{}^{\beta}\,\Gamma_{\bot}{}^{\gamma\alpha}
\qquad\Longleftrightarrow\qquad\ell_{n}\underline{R}^{\alpha \beta}\equiv
\underline{D}(\ell_{n}\underline{\Gamma}^{\alpha \beta})\, ,
\eqno(\z)
$$
and
$$
\underline{D}\;\underline{R}^{\alpha \beta}\equiv 0.\eqno(\z)
$$
Using the  operator
$$
\gamma^{\alpha}{}_{\beta}:= \delta^{\alpha}_{\beta} -
\widetilde{n}^{\alpha}\,\widetilde{n}_{\beta}\eqno(\z)
$$
the spatial components $A, B, C,\cdots=1,2,3$ of the above expressions
can be readily projected out.
\bigskip
\centerline {\bf Acknowledgments}
We cordially thank Romulado Tresguerres (Madrid) for useful
suggestions.  One of us (E.W.M.) wishes to acknowledge vivid discussions
with Alfredo Mac\'\i as (UNAM, Mexico, D.C.) and support of Conacyt.
Research support for Y.N.O. was provided by the Alexander von Humboldt
Foundation (Bonn).

\goodbreak
\bigskip
\centerline {\bf References}
\newref
[1] F.A. Lunev, {\sl Phys. Lett.} {\bf B295} (1992) 99.
\newref
[2] F.A. Lunev, {\sl Theor. and Math. Phys.} {\bf 94} (1993) 48.
\newref
[3] F.A. Lunev, {\sl Phys. Lett.} {\bf B311} (1993) 273, {\bf B314}
(1993) 21.
\newref
[4] K. Johnson, {\it The Yang--Mills ground state}, in: {\sl QCD -- 20
Years Later, Aachen 1992}, P.M. Zerwas and H.A. Kastrup, eds., Vol.2
(World Scientific, Singapore 1993) p.795.
\newref
[5] D.Z. Freedman, P.E. Haagensen, K. Johnson, and J.I. Latorre, {\it The
hidden spatial geometry of non--Abelian gauge theories}, {\sl preprint
CERN-TH.7010/93} (1993), submitted to Nucl. Phys. B.
\newref
[6] D.Z. Freedman, and R.R. Khuri, {\it The Yang--Mills ambiguities
revisited}, {\sl Preprint CERN-TH.7187} (March 1994); hep-th 9403031.
\newref
[7] M. Bauer, D.Z. Freedman, and P.E. Haagensen, {\it Spatial geometry
of the electric field representation of non--abelian gauge theories},
{\sl preprint CERN-TH.7238/94} (April 1994); hep-th 9405028.
\newref
[8] Yu.N. Obukhov, F.W. Hehl, and E.W. Mielke, {\it On the relation
between internal and external gauge theories}, {\sl preprint (Inst.
Theor. Phys., Univ. of Cologne: March 1994)}.
\newref
[9] Y. Ne'eman and Dj. \v{S}ija\v{c}ki, {\it Chromogravity: QCD-induced
diffeomorphisms}, {\sl Preprint Tel--Aviv University} (1993).
\newref
[10] Dj. \v{S}ija\v{c}ki, {\it QCD originated dynamical symmetry for
hadrons}, {\sl Preprint hep-th 9311033} (November 1993).
\newref
[11] Abdus Salam and C. Sivaram, {\sl Mod. Phys. Lett.} {\bf A8}
(1993) 321.
\newref
[12] V. Husain and K.V. Kuchar, {\sl Phys. Rev.} {\bf D42} (1990)
4070.
\newref
[13] E.W. Mielke, {\sl Ann. Phys. (N.Y.)} {\bf 219} (1992) 78.
\newref
[14] F.W. Hehl, {\it On the axiomatics of Maxwell's theory}. Lecture
given at the Annual Meeting of the German Physical Society (DPG),
Hamburg, March 1994.
\newref
[15] F.W. Hehl, J. Lemke, and E.W. Mielke, {\it Two lectures on fermions
and gravity}, in: Proc. of the School on
{\sl Geometry and Theoretical Physics}, Bad Honnef, 12 --16 Feb. 1990,
J. Debrus and A.C. Hirshfeld, eds. (Springer, Berlin 1991), p.56.
\newref
[16] J.D. Bjorken and S. Drell, {\it Relativistic Quantum Mechanics}
(Mc Graw--Hill, New York 1964).
\newref
[17] E.W. Mielke, in: {\it Differential Geometric Methods in
Mathematical Physics}, Proc. of the Int. Conference held at the
Technical University of Clausthal, Germany, July 1978, {\sl Lecture
Notes in Physics}, Vol. {\bf 139}, ed. H.D. Doebner (Springer, Berlin
1981), p.135.
\newref
[18] E.W. Mielke, {\it Geometrodynamics of Gauge Fields ---
On the geometry of Yang--Mills and gravitational gauge theories}
(Akademie--Verlag, Berlin 1987).
\newref
[19] S. Gotzes and A.C. Hirshfeld, {\sl Ann. Phys. (N.Y.)} {\bf 203}
(1990) 410.
\newref
[20] F.W. Hehl and J.D. McCrea, {\sl Found. Phys.} {\bf 16} (1986) 267.
\newref
[21] R.D. Hecht, {\it Erhaltungsgr\"o\ss en in der Poincar\'e Eichtheorie
der Gravitation}, {\sl Ph.D. thesis} (University of Cologne 1993).
\newref
[22] J. Schwinger, {\sl Phys. Rev.} {\bf 130} (1963) 1253.
\newref
[23] E.W. Mielke and R.P. Wallner, {\sl Nuovo Cimento} {\bf 101B}
(1988) 607; {\bf 102B} (1988) 555 (E).
\newref
[24] E.W. Mielke and P. Baekler, {\sl Phys. Lett.} {\bf A156}, 399
(1991); P. Baekler, E.W. Mielke, and F.W. Hehl, {\sl Nuovo Cimento}
{\bf B107} (1992) 91.
\newref
[25] B.P. Dolan, {\sl J. Phys.: Math. and Gen.} {\bf A15} (1982) 2191.
\newref
[26] A. Ashtekar, {\sl Phys. Rev. Lett.} {\bf 57} (1986) 2244; {\it
New Perspectives in Canonical Gravity} (Bibliopolis, Napoli 1988).
\newref
[27] E.W. Mielke, {\sl Phys. Rev.} {\bf D42} (1990) 3388.
\newref
[28] J.M. Nester, R.--S. Tung, and Y.Z. Zhang, {\sl Class. Quantum
Grav.} {\bf 11} (1994) 757.


\vfill\bye